\documentclass[prl,twocolumn,showpacs,floatfix,amsmath,amssymb, A4,superscriptaddress]{revtex4-1}

\usepackage{hyperref}
\hypersetup{colorlinks=true, citecolor=blue, urlcolor=blue, linkcolor=blue}
\usepackage{color}

\usepackage{colordvi}
\usepackage{amssymb}
\usepackage{amsmath}
\usepackage{epsf}
\usepackage{mathrsfs}
\usepackage{graphicx}
\usepackage{appendix}

\def\nn{\nonumber}
\def\be{\begin{equation}}
\def\ee{\end{equation}}
\def\bea{\begin{eqnarray}}
\def\eea{\end{eqnarray}}
\begin{document}

\title{\textbf{Self-bound supersolid stripe phase in binary Bose-Einstein condensates}}

\author{R.~Sachdeva}
\email{rashi.sachdeva@matfys.lth.se}
\affiliation{Mathematical Physics and NanoLund, LTH, Lund University, Box 118, 22100 Lund, Sweden}

\author{$\!\!^{,}\,^{\dagger}\,\,\,$M.~Nilsson Tengstrand}
\thanks{These two authors contributed equally.}
\affiliation{Mathematical Physics and NanoLund, LTH, Lund University, Box 118, 22100 Lund, Sweden}

\author{S.~M.~Reimann}
\affiliation{Mathematical Physics and NanoLund, LTH, Lund University, Box 118, 22100 Lund, Sweden}

\date{\today}

\begin{abstract}
Supersolidity -- a coexistence of superfluidity and 
crystalline or amorphous density variations -- has been vividly debated ever since its conjecture. 
While the initial focus 
was on helium-4, recent experiments uncovered supersolidity in ultra-cold dipolar quantum gases.  
Here, we propose a new self-bound supersolid phase in a binary mixture of Bose gases with short-range interactions, making use of the non-trivial properties of spin-orbit coupling. 
We find that a first-order phase transition from a self-bound supersolid stripe phase to a zero-minimum droplet state of the Bose gas
occurs as a function of the Rabi coupling strength. These phases are characterized using the momentum distribution, the transverse spin polarization and the superfluid fraction. The critical point of the transition is estimated in an analytical framework. The predicted density-modulated supersolid stripe and zero-minimum droplet phase should be experimentally observable in a binary mixture of $^{39}$K with spin-orbit coupling. 
\end{abstract}

\maketitle
  

The formation of self-bound liquid droplets results from a balance of effective attractive and repulsive forces between their constituent particles. Macroscopic examples such as water or oil drops are ubiquitous.  In the quantum realm such self-bound liquid droplets are well-known to occur for dense quantum liquids such as atomic nuclei~\cite{BohrMottelson} or liquid helium~\cite{donelly1991,toennies2004,ancilotto2018}. Recently, ultracold atomic gases have emerged as versatile candidates where stable  droplets~\cite{bulgac2002,petrov2015,petrov2016} can be realized even in the very dilute quantum regime. Such self-bound states were found to form in dipolar Bose Einstein condensates (BEC) of dysprosium~\cite{kadau2016,schmitt2016,ferrierbarbut2016,ferrierbarbut2018} or erbium \cite{chomaz2016}. They may also form out of binary BEC mixtures of potassium in different hyperfine states~\cite{cabrera2018,semeghini2018}, as originally predicted in Refs.~\cite{petrov2015,petrov2016}. These ultra-cold atomic quantum droplets are bound and stabilized by balancing a residual mean field attraction with a weak repulsion originating from first-order contributions to the energy beyond mean field (BMF), usually referred to as the Lee-Huang-Yang (LHY) correction~\cite{lhy1957}. This new form of self-bound ultra-dilute quantum matter has already revealed a range of unexpected phenomena. 
Prominent examples are the pattern formation of quantum ferrofluidic droplets \cite{kadau2016,schmitt2016,ferrierbarbut2016} shaped by the long-range anisotropic dipolar interaction, or  the recent evidence for supersolidity~\cite{tanzi2019,bottcher2019,chomaz2019,Tanzi_arxiv_Dec2019,roccuzzo2019,roccuzzo2019rotating}. 

The supersolid (SS) phase of matter, characterized by the counterintuitive coexistence of solid and friction-free superfluid behavior,
has been long-sought and debated  \cite{leggett1970,balibar2010,boninsegni2012,chan2013,pomeau1994}, with helium-4 being a prime candidate for its observation. In connection with ultra-cold atomic gases supersolidity has been discussed for trapped BEC with soft-core two-body potentials~\cite{pomeau1994,henkel2010,cinti2010,saccani2011},
for a BEC coupled to two optical cavities~\cite{leonard2017,leonard2017_nature}, and notably, also for trapped BECs with spin-orbit coupling (SOC)~\cite{Zhai_PRL2010,lin2011,y_li2012,JWPan_PRL2012,Engels_PRA2013,Stringari_superstripes_PRL2013,j_li2016,Wu_83}. 

In dipolar systems a phase transition from the single quantum droplet phase to a density-modulated SS phase has recently been observed~\cite{tanzi2019,bottcher2019,chomaz2019,Tanzi_arxiv_Dec2019}. The transition here results from the interplay between long-range anisotropic and short-range isotropic interactions with quantum fluctuations, as discussed above, with its mechanism now being well understood. Subsequent works~\cite{tanzi2019nature,guo2019,natale2019} have demonstrated the measurement of the Goldstone modes depicting the spontaneous symmetry breaking related to the SS phase. In a spin-orbit coupled BEC~\cite{Zhai_PRL2010,lin2011,y_li2012,JWPan_PRL2012,Engels_PRA2013,Stringari_superstripes_PRL2013,j_li2016,Wu_83}, the formation of density modulations in the SS phase appear as a result of momentum transfer achieved through the Raman coupling between pseudospin states formed by two atomic hyperfine ground states. Hence, the physical mechanism for the occurrence of supersolidity for the spin-orbit coupled system is very different from that in dipolar condensates.

An intriguing question is
whether such a SS phase can also exist in {\it self-bound} binary droplets with short-range interactions,  
for example made from potassium~\cite{cabrera2018,semeghini2018}.
In this Letter, employing the non-trivial 
properties of SOC~\cite{Zhai_PRL2010,lin2011,y_li2012,JWPan_PRL2012,Engels_PRA2013,Stringari_superstripes_PRL2013,j_li2016,Wu_83}, we find that a self-bound density-modulated SS stripe phase is induced in  the ground state of a LHY-stabilized binary BEC. With increasing Rabi coupling strength the self bound SS stripe phase changes to a zero-minimum (ZM) droplet phase by a first-order phase transition. We employ a numerical extended Gross-Pitaevskii (eGP) approach~\cite{baillie2016,wachtler2016}, and corroborate our findings with an analytical variational approach.

Let us now consider a two-dimensional weakly interacting binary BEC, as in~\cite{petrov2016}, but here with a Raman-induced equal 
Rashba-Dresselhaus SOC
\cite{lin2011,j_li2016,y_li2012} along the $x$-direction. The dimensionless energy functional with the modified interaction term \cite{petrov2015,petrov2016} describing this mixture is given by
\bea 
E&=&\int d\mathbf{r}\bigg[\bigg(\psi_{1}^{*}\frac{(p_x-\gamma)^2+p_y^2}{2}\psi_1\bigg)\nn\\
& &+\bigg(\psi_{2}^{*}\frac{(p_x+\gamma)^2+p_y^2}{2}\psi_2\bigg)+\frac{\Omega}{2}\big(\psi_{1}^{*}\psi_2+\psi_{2}^{*}\psi_1\big)\nn\\
& & +\frac{\delta}{2}\big(|\psi_1|^2-|\psi_2|^2\big)+\frac{g}{2}\big(|\psi_1|^2-|\psi_2|^2\big)^2\nn\\ & &+\frac{g^2}{8\pi}\big(|\psi_1|^2+|\psi_2|^2\big)^2~\text{ln}\frac{|\psi_1|^2+|\psi_2|^2}{\sqrt{e}}\bigg].\label{dropletenergyfunc}
\eea
\begin{figure}[tb]
\centering
\includegraphics[width=\columnwidth]{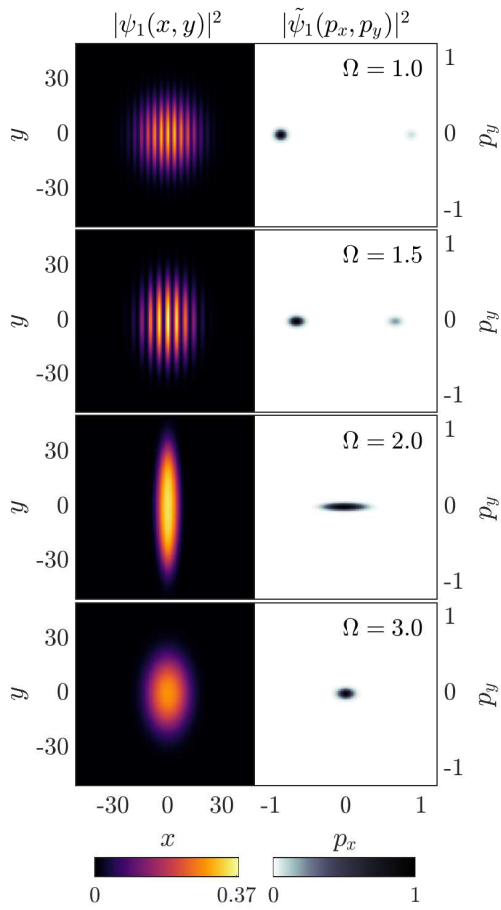}
\caption{\textit{(Color online)} Density profiles $|\psi_1(x,y)|^2$ {\it (left panels)} and the corresponding momentum distribution profiles $|\widetilde{\psi}_1(p_x,p_y)|^2$ {\it (right panels)} (normalized such that its maximum value is unity) of the first component for varying values of dimensionless Rabi coupling strength $\Omega=(1,1.5,2,3)$  for a fixed coupling constant $g=0.5$, SOC strength $\gamma=1$ and norm $N=400$.}
\label{Fig:1}
\end{figure}
Here $\psi_1$ and $\psi_2$ are the wave functions for the two components of the binary mixture, $\gamma$ is the SOC strength, $\Omega$ is the Rabi coupling strength, $\delta$ is the detuning and $g>0$ is the two-dimensional coupling constant. All interactions are short-range, and intra- and interspecies interactions are repulsive and attractive, respectively, implying $(g_{11},g_{22})>0$ and $g_{12}<0$. We take them to be pseudospin-symmetric, {\it i.e.}, $g_{11}=g_{22}=g$, and work in the stable regime~\cite{petrov2015,petrov2016}. According to Ref.~\cite{petrov2016}, the expression for the LHY energy can be written in a simpler form as in Eq.~\eqref{dropletenergyfunc}, provided the condition $n\ll a^{-2}$ is satisfied, where $n$ is the peak value of the two-dimensional atomic density and $a$ is the two-dimensional scattering length.
The detuning $\delta$ can be set to zero with a proper choice of the frequency of the two Raman lasers, hence we assume $\delta=0$. As suggested in Refs.~\cite{Zheng_2013,Malomed_PRA_2017}, the presence of moderately strong SOC in the binary BEC mixture can affect the BMF correction to the mean field energy resulting in an effective renormalization of the strength of the original mean field and the LHY correction. However, we work in a regime where the strength of the SOC is relatively weak and does not qualitatively affect the form of the mean field and BMF terms. We thus adopt the usual form of the mean field and BMF interaction terms in two dimensions~\cite{petrov2015,petrov2016}.

To determine the many-body ground state of the two components, we obtain the coupled eGP equations through the standard minimization procedure of the energy functional in Eq.~\eqref{dropletenergyfunc} with respect to $\psi_1^\ast$ and $\psi_2^\ast$. 
The resulting dimensionless eGP equations are solved through the usual split-step Fourier method using the imaginary time propagation technique. We  use a large set of different initial conditions in order to avoid local minima in the energy. 
Figure~\ref{Fig:1} shows the eGP density profiles and the corresponding momentum distribution of the first component for fixed SOC strength $\gamma=1$, fixed coupling constant $g=0.5$ and varying values of the Rabi coupling strength $\Omega=(1,1.5,2,3)$. The second component has identical ground state density distributions and mirrored momentum distribution profiles and is thus not shown in Fig.~\ref{Fig:1}.

Intriguingly, the formation of a self-bound SS stripe phase can already be observed for low values of $\Omega$, as depicted in Fig.~\ref{Fig:1}. These density modulations realized in the SOC system appear as a result of coupling two pseudospin states of the BEC by Raman lasers, modifying the single-particle dispersion to possess a minimum at finite momentum values. Unlike for dipolar condensates, where the density modulations in the SS phase arise from the long-ranged interactions, the interplay between the modified single-particle dispersion and the two-body interactions is the source of the supersolidity in the present case. Since the SOC here is chosen to act only along $x$-direction, the formation of stripes is only observed in this direction. As mentioned above, in the eGP the attractive Bose gas is stabilized by the LHY correction~\cite{lhy1957}, which renders the SS system self-bound, in analogy to the dipolar case albeit here 
solely with contact interactions.   
For higher values of $\Omega$ we find a phase transition from a self-bound SS stripe (with non-zero canonical momentum) to a self-bound ZM droplet (zero canonical momentum).

Let us now proceed with an analytical approach in order to further analyze the interplay between the non-trivial properties of the single-particle dispersion relation of the spin-orbit coupled BEC, and the additional BMF corrections. 
Following~\cite{y_li2012}, we make the following initial ansatz for the two component wave function:
\bea \psi_{1}&=&\sqrt{\bar{n}}\big[C_1~\text{cos}~\theta~e^{ik_1x}+C_2~\text{sin}~\theta~e^{-ik_1x}\big]\nn\\
\psi_{2}&=&-\sqrt{\bar{n}}\big[C_1~\text{sin}~\theta~e^{ik_1x}+C_2~\text{cos}~\theta~e^{-ik_1x}\big].\label{ansatz}
\eea
Here, $\bar{n}=N/V$ is the average density and $k_1$ is the canonical momentum. The variational parameters in the present case are $C_1, C_2, k_1$ and $\theta$. They are determined by minimizing Eq. \eqref{dropletenergyfunc} along with the normalization constraint $\int d\mathbf{r}~(|\psi_1|^2+|\psi_2|^2)=N$, implying $|C_1|^2+|C_2|^2=1$. Using the single particle Hamiltonian, the minimization with respect to $\theta$ gives the general relation $\theta=\text{arccos}(k_1/\gamma)/2$. 

The inclusion of both mean field and BMF interactions will affect the variational parameters $C_1$, $C_2$ and $k_1$. We use the minimization of the energy functional  Eq.~\eqref{dropletenergyfunc} to determine these parameters. Substituting Eq.~\eqref{ansatz} into Eq.~\eqref{dropletenergyfunc}, we simplify the energy per particle $\epsilon=E/N$ as
\bea \epsilon&=&\frac{\gamma^2}{2}-\frac{\Omega}{2\gamma}\sqrt{\gamma^2-k_1^2}-\frac{k_1^2}{2\gamma^2}F(\beta)\nn\\
& & +F_{LHY}\bigg[1+2\beta\bigg(1-\frac{k_1^2}{\gamma^2}\bigg)\bigg],\label{dropsubstotalenergy}\eea
where $\beta=|C_1|^2|C_2|^2$ is a dimensionless parameter that can assume values $0\leq\beta\leq 1/4$ and $F_{LHY}=(\bar{n}g^2/8\pi)~\text{ln}(\bar{n}/\sqrt{e})$. We also introduced the function $F(\beta)=\gamma^2+2G_2(4\beta-1)$ with the interaction parameter $G_2=\bar{n}(g-g_{12})/4$. Eq.~\eqref{dropsubstotalenergy} shows the modified version of the energy per particle for a uniform spin-orbit coupled BEC due to the presence of the stabilizing LHY terms beyond mean field. The last term in Eq.~\eqref{dropsubstotalenergy} arises solely because of the inclusion of the BMF effects,  essentially modifying the properties of the system compared to the usual spin-orbit coupled BEC. The condition  
$g_{11}=g_{22}=g$ and $g_{12}=-g$ implies $G_1=\bar{n}(g+g_{12})/4=0$ and $G_2=\bar{n}(g-g_{12})/4=\bar{n}g/2\neq 0$, hence the expression for $F(\beta)$ in our case can be seen as a limiting case of $F(\beta)$ for the usual spin-orbit coupled BEC~\cite{y_li2012}. 
Note that the contributions from the BMF effect can be simplified as a sum of two terms, one of which appears as the last term in Eq.~\eqref{dropsubstotalenergy}. We drop the second contribution from the BMF effect in Eq.~\eqref{dropsubstotalenergy} because it was found to be negligible compared to the other terms for the extreme cases i)~$\beta=1/4$, $k_1\neq 0$, and ii)~$\beta=0,k_1=0$. To proceed further, we first perform the minimization with respect to $k_1$ by using $\partial\epsilon/\partial k_1=0$, yielding
\be k_1(\beta)=\gamma\sqrt{1-\frac{\Omega^2}{4[H(\beta)]^2}},\label{dropk1}\ee
where $H(\beta)$ is defined as $ H(\beta)=F(\beta)+4\beta F_{LHY}$.
Eq.~\eqref{dropk1} minimizes the energy for $\Omega<2H(\beta)$. 
The comparison of $H(\beta)$  with the mean field results~\cite{y_li2012} explicitly shows the modification of the term $F(\beta)$ beyond mean field, leading to a change in the value of the canonical momentum $k_1(\beta)$ where the BEC takes place.

Further, we substitute Eq.~\eqref{dropk1} into Eq.~\eqref{dropsubstotalenergy} and obtain
\bea \epsilon&=& -\frac{\Omega^2}{4H(\beta)}+\frac{\Omega^2}{8[H(\beta)]^2}F(\beta)+G_2(1-4\beta)\nn\\
&&+F_{LHY}\bigg[1+\frac{\Omega^2\beta}{2[H(\beta)]^2}\bigg].\label{totalenergyfin}\eea
We can find the ground state  by looking for the minimum of Eq.~\eqref{totalenergyfin} with respect to $\beta$. 
The ground state turns out to appear in the following two phases. \textit{Self-bound supersolid (SS) stripe phase}: This phase is favored for small values of Raman coupling $\Omega$ with $|C_1|=|C_2|=1/\sqrt{2}$, resulting in density modulations in the form of stripes following
\bea n(\mathbf{r})&=&\bar{n}\bigg[1+\frac{\Omega}{2(\gamma^2+F_{LHY})}~\text{cos}~(2k_1x+\phi)\bigg].\label{densmod}\eea
In addition, the periodicity ($\pi/k_1$) of the modulations is calculated using the wave number
\bea k_1&=& \gamma\sqrt{1-\frac{\Omega^2}{4(\gamma^2+F_{LHY})^2}}.\eea 
These modulations appear as the result of spontaneous breaking of the translational invariance in the self-bound droplet phase. This has evident analogies with the supersolid phase in terms of the spontaneous formation of periodic density modulations. \textit{Self-bound zero-minimum (ZM) droplet phase}: For higher values of $\Omega $, the system undergoes a phase transition to the self-bound ZM droplet phase, where the atoms stay in the zero momentum phase ($k_1=0$) and the density is uniform for both components. The analytically obtained ground state phases are in good agreement with the numerical observations. In addition, we can also determine the critical value of Rabi coupling strength for the phase transition by using the respective energies of the ZM droplet phase ($k_1=0$) and the SS stripe phase ($\beta=1/4, k_1\neq 0$). The value of $k_1$ gets smaller as we move towards the phase transition point, and hence the energy of the SS stripe phase is further amended with an approximation to the previously neglected contribution to the energy (which is a good estimate when $k_1$ is close to zero at the phase transition). The critical value of Rabi coupling strength turns out to be
\bea \Omega_{c}&=& 2H(1/4)\bigg[1-\sqrt{1-\frac{\gamma^2-2I_2}{\gamma^2+F_{LHY}}}\bigg],\label{omegacric}\eea 
where $I_2=(\bar{n}g^2/16\pi)\big[7/2-3~\text{ln}~2\big]$.
In what follows below, we further characterize the phases by calculating the relevant physical observables, such as the canonical
momentum $k_1$ as given in Eq.~\eqref{dropk1}, and  
the transverse spin polarization $\langle\sigma_x\rangle=-\sqrt{1-k_1^2/\gamma^2}$, in order to quantify the phase transition between the SS stripe and ZM droplet phase.

\begin{figure}[tb]
\centering
\includegraphics[width=\columnwidth]{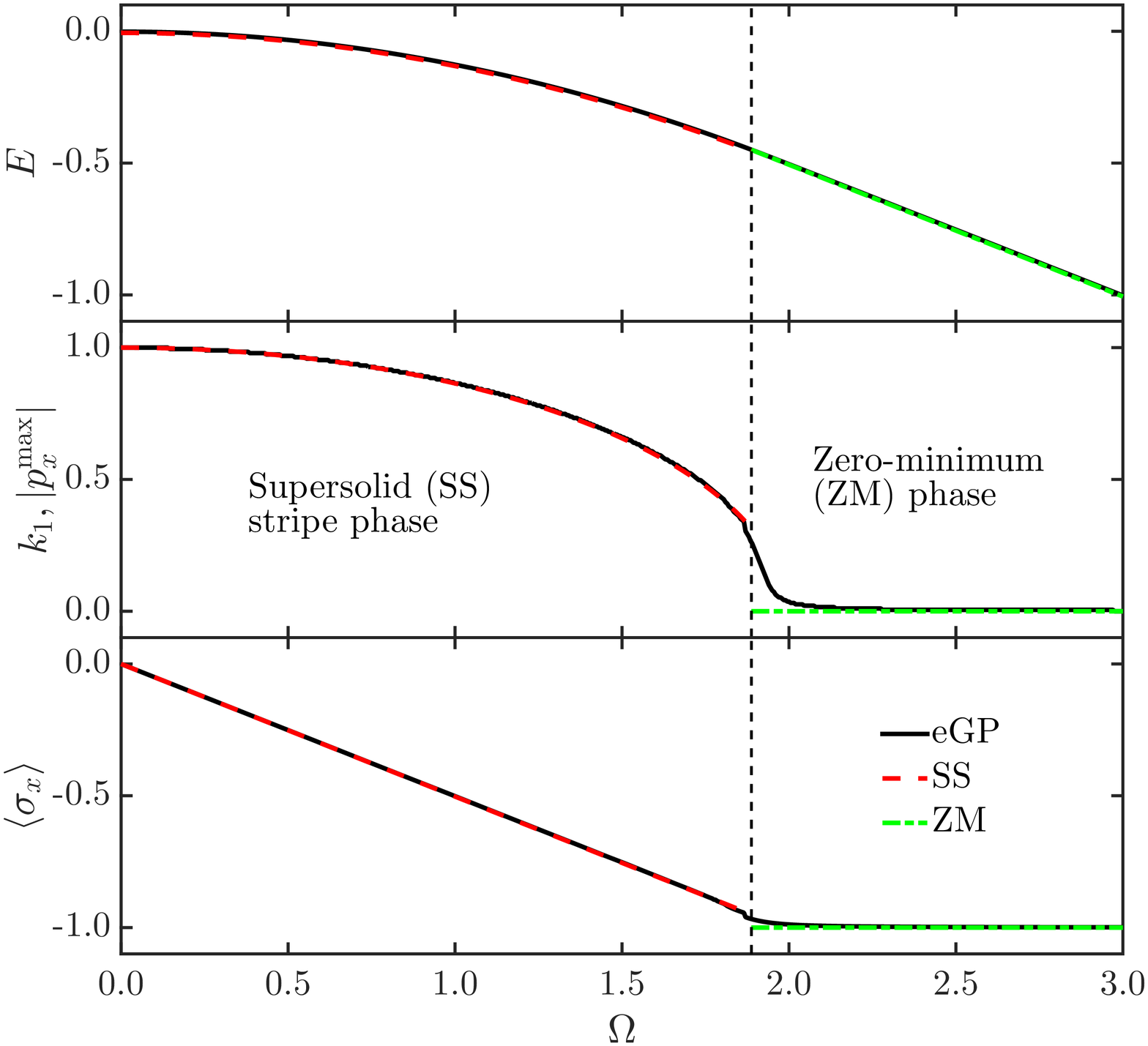}
\caption{\textit{(Color online)} Energy per particle $\epsilon=E/N$, canonical momentum $k_1$ and transverse polarization $\langle\sigma_x\rangle$ as a function of dimensionless Rabi coupling strength $\Omega$ for $g=0.5$, $\gamma=1$ and $\delta=0$. $|p_x^{max}|$ is the value for the highest peak in the momentum distribution in the $x$ direction. One can notice the discontinuity in $k_1$ and $\langle\sigma_x\rangle$ at the phase transition point $\Omega_c\approx1.89$. Red dashed line, green dashed-dotted lines and black solid lines depict the observables calculated analytically for the SS stripe phase $k_1\neq0$, $\beta=1/4$, for the ZM droplet phase $k_1=0$, and the numerical eGP results, respectively. The black dashed vertical line indicate the analytical estimate of the critical Rabi coupling strength for the phase transition. We use the equilibrium value of the density $\bar{n}=1/\sqrt{e}$ to compare the numerical and analytical results.}
\label{Fig:2}
\end{figure}

Using both the numerical  and analytical results, we plot the energies along with the momentum $k_1$ and transverse spin polarization $\langle\sigma_x\rangle$ in Fig.~\ref{Fig:2}. The plots clearly show a first-order phase transition as the value of Rabi coupling strength $\Omega $ is increased. This is depicted in Fig.~\ref{Fig:2} (b) and (c) by a jump in the value of $k_1$ and consequently a jump in $\langle\sigma_x\rangle$ as we transverse from the self-bound SS stripe to the ZM droplet regime. For our choice of parameters, the results from the numerical and analytical calculations agree well with each other, indicating the general applicability of the variational ansatz for the spinor wave function used to determine the ground states for SOC systems.

\begin{figure}[tb]
\centering
\includegraphics[width=\columnwidth]{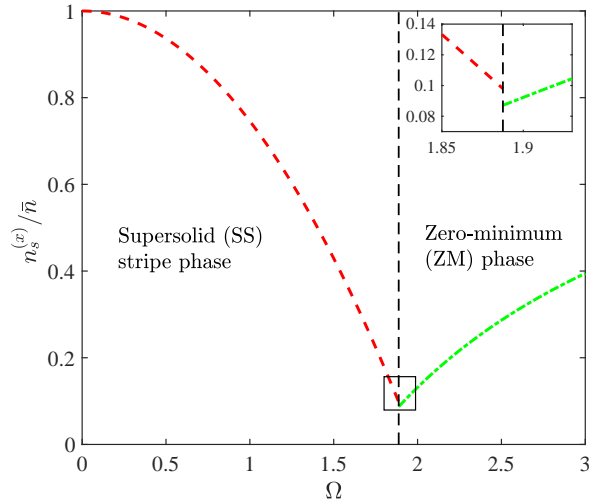}
\caption{\textit{(Color online)} Superfluid fraction characterized by superfluid density as a function of the Rabi coupling strength in the SS stripe phase (red dashed line) and the ZM droplet phase (green dashed-dotted line). The black dashed vertical line indicate the analytical estimate of the critical Rabi coupling strength for the phase transition. Other parameters are the same as in Fig.~\ref{Fig:2}.}
\label{Fig:3}
\end{figure}

We finally characterize the superfluidity of the system by calculating the superfluid density, which shows a unique behavior in the presence of SOC. We implement a phase
twist method~\cite{fisher1973} to calculate the superfluid density along the 
$x$ and $y$ directions using the analytical expressions of the energy for the SS stripe and ZM droplet phase. The behavior of the
superfluid density $n_s^{(x)}$ along the SOC direction as a function of the 
Rabi coupling strength is shown in Fig. \ref{Fig:3}. It exhibits an interesting behavior in the SS stripe and ZM regimes: it falls down monotonically in the
SS stripe phase, goes to a small value at the critical transition point and is followed by an increase in the ZM droplet phase. As expected, the superfluid density shows a
discontinuity at $\Omega_c$ which is also a signature of a first order phase
transition between the SS stripe and the ZM droplet phase. There is a suppression in the superfluid density for small increasing values of $\Omega$ because of the crucial role played by the breaking of
Galilean invariance and the gapped branch of the elementary excitations~\cite{stringari2016} in the SOC system.
Note that there is no density modulation in the $y$ direction due to the 
absence of SOC, hence the superfluid fraction is $n_s^{(y)}/\bar{n}=1$ as in a conventional BEC.

Until now, SOC has been realized in experiments in $^{87}$Rb bosonic species. However recent experimental progress \cite{sanz2019interaction} with coupling of two internal states ($|F,m_F\rangle=|1,-1\rangle$ and $|1,0\rangle$) of $^{39}$K  resulting in modification of the scattering properties of the corresponding dressed states is a promising step to achieve SOC in binary BEC mixtures in the self-bound regime.

In conclusion, we have studied the realization of a self-bound supersolid in a binary Bose-Einstein condensate mixture using the non trivial properties of spin-orbit coupling, and its phase transition to a zero-minimum droplet phase. We applied a numerical extended Gross-Pitaevskii approach and compared it to an analytical variational calculation to determine the ground state quantum phases characterized by the canonical momentum, the transverse spin polarization and the superfluid density. The findings presented here are in the range of the current state of the art of experiments for binary condensates. The present analysis opens up new perspectives for spin-orbit coupling and supersolid phenomena in self-bound quantum systems.\\
\textit{Acknowledgements}: We thank the Knut and Alice Wallenberg Foundation and the Swedish Research Council for financial support.
 

\bibliography{SOC_droplet_final.bib}

\end{document}